\documentclass{ifacconf}

\usepackage{graphicx}      
\usepackage{natbib}        
\usepackage{amssymb}
\usepackage{lipsum}
\usepackage{amsmath}
\usepackage{gensymb}
\usepackage{float}
\usepackage{graphicx}
\usepackage{subcaption}
\usepackage{tabularx}
\usepackage{comment}
\usepackage{dblfloatfix}
\usepackage{xcolor}
\definecolor{mycolor1}{RGB}{0 0.4470 0.7410}
\definecolor{mycolor2}{RGB}{0.70,0.00,0.30}
\usepackage{wrapfig}
\usepackage{soul}
\usepackage{multirow}
\usepackage{cuted}
\usepackage{tikz}
\begin{document}
\begin{frontmatter}

\title{Structured $\mu$-Synthesis for Nanopositioners under Payload-Induced Uncertainties: Minimising Conservatism for Robust Performance} 

\thanks[footnoteinfo]{This work was financed by Physik Instrumente (PI) SE \& Co. KG and co-financed by Holland High Tech with PPS Project supplement for research and development in the field of High Tech Systems and Materials.}

\author[First]{Manavi Araga} 
\author[First]{Aditya Natu} 
\author[First]{Hassan HosseinNia} 

\address[First]{Department of Precision and Microsystems Engineering, Delft University of Technology, Mekelweg 2, 2628 CD Delft,
The Netherlands (e-mail:  a.m.natu@tudelft.nl; s.h.hosseinniakani@tudelft.nl)}

\begin{abstract}
Most systems exhibit significant variability in their dynamics, including variations in system parameters and large high-frequency dynamic uncertainties. Traditional uncertainty modelling techniques consolidate all such variations into a single uncertainty block, often yielding overly conservative representations of the true plant behaviour. This paper introduces an uncertainty modelling framework that employs multiple structured and unstructured uncertainty blocks to reduce this conservatism. The methodology is evaluated for an industrial piezoelectric nanopositioner subject to payload-induced variations, using uncertainty models of differing complexity. A bandpass controller is synthesised via structured mixed-$\mu$ synthesis, and the resulting designs are compared in terms of conservatism of the uncertainty model, robust performance, and computational effort.
\end{abstract}

\begin{keyword} Uncertainty Modelling \sep Structured Uncertainty \sep Unstructured Uncertainty \sep Structured Mixed-$\mu$ Synthesis\sep Payload-Induced Uncertainties \sep Piezoelectric Nanopositioner
\end{keyword}

\end{frontmatter}

\section{Introduction}
\label{sec:Introduction}
Accurate plant modelling is fundamental to effective control design. Although control design procedures typically assume that the plant model perfectly represents the actual system, obtaining an exact representation of a physical system across the entire frequency range of interest is generally infeasible with linear identification methods. At high frequencies, where the system has low gain, the identified dynamics become unreliable due to noise contamination (\cite{pintelon2012system}). Measurement imperfections and dependency on operating conditions further affect the identified parameters. Nonlinearities such as hysteresis introduce amplitude-dependent variations, and although inverse hysteresis approaches mitigate these effects (\cite{al2023prandtl}), they are not exact. In numerous precision motion systems, conditions such as mass loading shift the resonant modes (\cite{natu2025robust}). Consequently, the identified parameters remain approximations of the true system behaviour and may vary across operating conditions.

Controller designs that ignore such variations may experience degraded performance or even instability when deployed on the real system. To mitigate this, robust control methods incorporate a set of plant models representing the expected variations into the control synthesis process. Uncertainty modelling techniques are used to construct an uncertain plant model that captures the anticipated variations. Control system performance is then evaluated with respect to this uncertain plant model. However, as the domain of uncertainty increases, both robust stability and robust performance generally deteriorate (\cite{yedavalli2014robust}). It is therefore essential that the uncertain plant model closely reflects the measured or estimated parameter variations without unnecessary conservatism.

Uncertainties primarily originate from two sources: (i) variations in physical parameters governing plant behaviour, such as resonant frequencies and damping ratios, and (ii) unmodelled or neglected dynamics, including higher-order resonances beyond the frequency range of interest or unknown system delays. The first category, known as parametric uncertainty, arises from measurable or predictable variations in system parameters. The second category, referred to as dynamic uncertainty, results from high-frequency dynamics that are either intentionally omitted to simplify analyses or unmodelled due to measurement limitations. Parametric uncertainties correspond to real-valued parameter variations and are typically represented using structured uncertainty blocks. In contrast, dynamic uncertainties may be real- or complex-valued and are represented using an unstructured uncertainty block (\cite{gu2005robust}).

In systems where plant dynamics are well-characterized within the frequency range of interest, uncertainties primarily stem from parameter variations and can be represented using a single structured uncertainty block. For example, \cite{shao2025design} considers parametric uncertainties arising from the moment of inertia, joint stiffness, and damping, while \cite{ahmad2020robust} assumes a 5\% variation in characteristic equation coefficients due to changes in operating conditions, such as temperature, humidity, etc. These variations are consolidated into a single structured uncertainty block. When dynamic uncertainties are also significant, both parametric and dynamic effects may be combined into a single unstructured uncertainty block (\cite{van2002multivariable, duan2024modeling}). Alternatively, \cite{gaspar2002robust} proposes a diagonal $2\times2$ uncertainty matrix, where the first diagonal element represents structured uncertainty encompassing all parametric variations, and the second diagonal element accounts for dynamic uncertainty. While these approaches adequately represent system uncertainties, the weighted formulation for unstructured uncertainties introduces notable conservatism (\cite{skogestad2005multivariable}). This conservatism may be further compounded when large, multi-parametric variations and dynamic uncertainties are combined into a single unstructured uncertainty block.

To address this, a methodology for uncertainty modelling that employs multiple uncertainty blocks is proposed to reduce conservatism and capture the large, multi-parametric variations. Since large uncertainty models restrict the domain of stabilizing controllers and degrade achievable performance, a minimally conservative uncertainty representation preserves a larger solution space, enabling synthesis results with improved performance. The effectiveness of the method is demonstrated by synthesizing a fixed-structure bandpass controller using structured mixed-$\mu$ synthesis for an industrial piezoelectric nanopositioning system, enhancing damping performance. Measured frequency response functions (FRFs) from the system are used to construct uncertainty models of varying complexity, depending on the level of conservatism, and the resulting robust performance is evaluated through simulation.

The remainder of this paper is organized as follows: Section \ref{sec:uncert} describes the proposed uncertainty modelling framework. Section \ref{sec:ex} details the validation of the methodology on an industrial piezoelectric nanopositioning system. Section \ref{sec:conclusion} concludes the paper and discusses potential future extensions of this work.

\section{Uncertainty Modelling}
\label{sec:uncert}
This section outlines the uncertainty modelling methodology, distinguishing between parametric and dynamic uncertainties. A mathematical framework for deriving the uncertainty weights associated with the structured and unstructured uncertainty blocks is described in this section.
\subsection{Structured Uncertainty}
Most flexible systems can be characterized as a combination of resonant and anti-resonant modes. The transfer function for the $j^{th}$ resonance and anti-resonance pair is:
\begin{equation}
g_j(s)=\cfrac{\left(s/Z_j\right)^2+2\zeta_{Z_j}s/Z_{j}+1}{\left(s/P_j\right)^2+2\zeta_{P_j}s/P_{j}+1}=\cfrac{n_{2j}s^2+n_{1j}s+1}{d_{2j}s^2+d_{1j}s+1},
\end{equation}
\begin{figure}[!t]
    \centering
    \includegraphics[width =\linewidth]{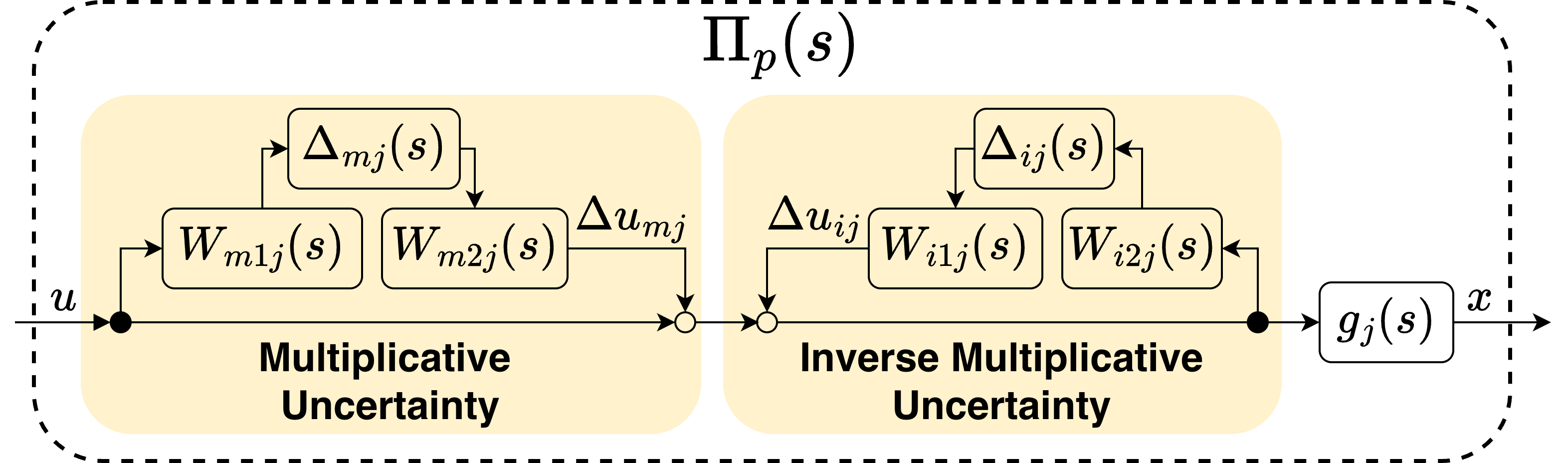}
    \caption{Structured uncertainty model for a resonance and anti-resonance pair.}
    \label{fig:unc_plant}
\end{figure}
where \( P_{j} \) and \( \zeta_{P_j} \) signify the frequency and damping ratio of the $j^{th}$ resonance, and \( Z_{j} \) and \( \zeta_{Z_j} \) denote the frequency and damping ratio of the $j^{th}$ anti-resonance. The uncertainties in the frequency or damping ratio of the $j^{th}$ resonance and anti-resonance pair are directly related to uncertainties in the coefficients of the numerator and denominator, denoted by \( n_{2j} \), \( n_{1j} \), \( d_{2j} \), and \( d_{1j} \). Uncertainty in the numerator coefficients, which affects the system zeros, can be represented using an input multiplicative uncertainty structure. Conversely, uncertainty in the denominator coefficients, which leads to variations in the system poles, is captured using an input inverse-multiplicative uncertainty structure, as illustrated in Fig.~\ref{fig:unc_plant} (\cite{zhou1998essentials}). Using this uncertainty structure, the uncertainty model for the $j^{\text{th}}$ resonance and anti-resonance pair, denoted by \( \Pi_p(s) \), is defined as follows:
\begin{equation} \label{eqn:1}
\begin{split}
    \Pi_p(s)=g_{j}(s)&\left(1-W_{i1j}(s)\Delta_{ij}(s)W_{i2j}(s)\right)^{-1}\\&\left(1+W_{m1j}(s)\Delta_{mj}(s)W_{m2j}(s)\right)
\end{split}
\end{equation}
where \( W_{m1j}(s) \) and \( W_{m2j}(s) \) denote the multiplicative uncertainty weights that capture uncertainties in \( n_{2j} \) and \( n_{1j} \), respectively, corresponding to uncertainty in the $j^{th}$ anti-resonance mode. \( W_{i1j}(s) \) and \( W_{i2j}(s) \) are the inverse multiplicative uncertainty weights corresponding to uncertainties in \( d_{2j} \) and \( d_{1j} \), respectively, associated with uncertainty in the $j^{th}$ resonance mode. 

Choosing a nominal transfer function, \(g_{j}(s)\), with the average parameter values \(\overline{n_2j}\), \(\overline{n_1j}\), \(\overline{d_2j}\), and \(\overline{d_1j}\) for the numerator and denominator coefficients, the uncertainty model \(\Pi_p(s)\) can be rewritten as:
\begin{equation}\label{eqn:2}
\resizebox{0.9\hsize}{!}{$
    \Pi_p(s)=\cfrac{\overline{n_{2j}}(1+{n_{2j}}_r\delta_{n2j}(s))s^2+\overline{n_{1j}}(1+{n_{1j}}_r\delta_{n1j}(s))s+1}{\overline{d_{2j}}(1+{d_{2j}}_r\delta_{d2j}(s))s^2+\overline{d_{1j}}(1+{d_{1j}}_r\delta_{d1j}(s))s+1}.$}
\end{equation}
Parameters of \eqref{eqn:2} are defined as: 
\begin{equation}
\resizebox{0.9\hsize}{!}{$
\begin{split}
    {n_{1j}}_r=\cfrac{max(n_{1j})-min(n_{1j})}{2\overline{n_{1j}}},~{n_{2j}}_r&=\cfrac{max(n_{2j})-min(n_{2j})}{2\overline{n_{2j}}},\\
    {d_{1j}}_r=\cfrac{max(d_{1j})-min(d_{1j})}{2\overline{d_{1j}}},~{d_{2j}}_r&=\cfrac{max(d_{2j})-min(d_{2j})}{2\overline{d_{2j}}}
\end{split}$}
\end{equation}
and \(||\delta_{n1j}(s)||_\infty\leq1\), \(||\delta_{n2j}(s)||_\infty\leq1\), \(||\delta_{d1j}(s)||_\infty\leq1\), \(||\delta_{d2j}(s)||_\infty\leq1\) are the real-valued, unity norm-bounded uncertainty blocks such that:
\begin{equation}\label{eqn:Delta_mult_invmult}
\resizebox{0.9\hsize}{!}{$
    \Delta_{mj}(s)=\begin{bmatrix}
	\delta_{n1j}(s) & 0\\0&\delta_{n2j}(s)
\end{bmatrix};~~\Delta_{ij}(s)=\begin{bmatrix}
	\delta_{d1j}(s) & 0\\0&\delta_{d2j}(s)
\end{bmatrix}.$}
\end{equation} 
Equating \eqref{eqn:1} and \eqref{eqn:2} and setting \(W_{m1j}(s)=W_{i1j}(s)=\begin{bmatrix}
	1&1
\end{bmatrix}\), the rest of the uncertainty weights are computed as:
\begin{align}
\resizebox{0.9\hsize}{!}{$
    W_{m2j}(s)=\begin{bmatrix}
		\cfrac{\overline{n_{1j}}{n_{1j}}_rs}{\overline{n_{2j}}s^2+\overline{n_{1j}}s+1}\\ \cfrac{\overline{n_{2j}}{n_{2j}}_rs^2}{\overline{n_{2j}}s^2+\overline{n_{1j}}s+1}
	\end{bmatrix};~ W_{i2j}(s)=\begin{bmatrix}
		\cfrac{-\overline{d_{1j}}{d_{1j}}_rs}{\overline{d_{2j}}s^2+\overline{d_{1j}}s+1} \\ \cfrac{-\overline{d_{2j}}{d_{2j}}_rs^2}{\overline{d_{2j}}s^2+\overline{d_{1j}}s+1}
	\end{bmatrix}.$}\label{eqn:weight_mult_invmult}
\end{align}
Depending on the number of parameters subject to uncertainty, \( W_{m2j}(s) \) and \( W_{i2j}(s) \) can be simplified. For instance, if uncertainty exists only in \( n_{2j} \) for $j=2$, the uncertainty model using solely the input multiplicative uncertainty structure, with the corresponding weight associated with \( n_{22} \), may be used.

For plant dynamics exhibiting multiple resonances and anti-resonances, several multiplicative and inverse multiplicative uncertainty blocks may be incorporated in series. To minimize conservatism in uncertainty modelling, a structured uncertainty block should be assigned to each uncertain parameter. However, increasing the number of uncertainty blocks increases the computational complexity of the synthesis problem. Consequently, an appropriate trade-off between conservatism reduction and computational effort is required.

\subsection{Unstructured Uncertainty}
Dynamic uncertainties are typically represented using an output multiplicative uncertainty structure. The uncertainty weight is designed to enclose the complex-valued relative multiplicative uncertainty \( E(s) \) defined as:
\begin{equation}
E(s) = \left|\frac{G_{e}(s) - G_{m}(s)}{G_{m}(s)}\right|.
\end{equation}
Here, $G_{e}(s)$ denotes the measured or estimated FRF of the system, while $G_{m}(s)$ represents the portion of the plant dynamics already captured by the 
structured uncertainty model. To reduce conservatism, the dynamic uncertainty weight $W_{d}(s)$, defined as $ E(s)\leq W_{d}(s)\Delta_{d}(s) $ for a 
complex-valued, unity norm-bounded uncertainty block $||\Delta_{d}(s)||_\infty\leq1$, is shaped to closely follow 
the contours of $E(s)$.

\section{Illustrative example}
\label{sec:ex}
The uncertainty modelling framework developed in Section \ref{sec:uncert} is applied to synthesise a bandpass controller aimed at improving damping at the first, dominant resonance of an industrial piezoelectric nanopositioning system with payload-induced uncertainties. This section outlines the system description, the associated uncertainty models, the chosen control architecture, the sensitivity weighting objectives and weighting function design, and the resulting performance.
\subsection{Piezoelectric Nanopositioning System Description}
\begin{figure}[t!]
\centering
\begin{subfigure}{\linewidth}
\centering
    \includegraphics[width=0.6\linewidth]{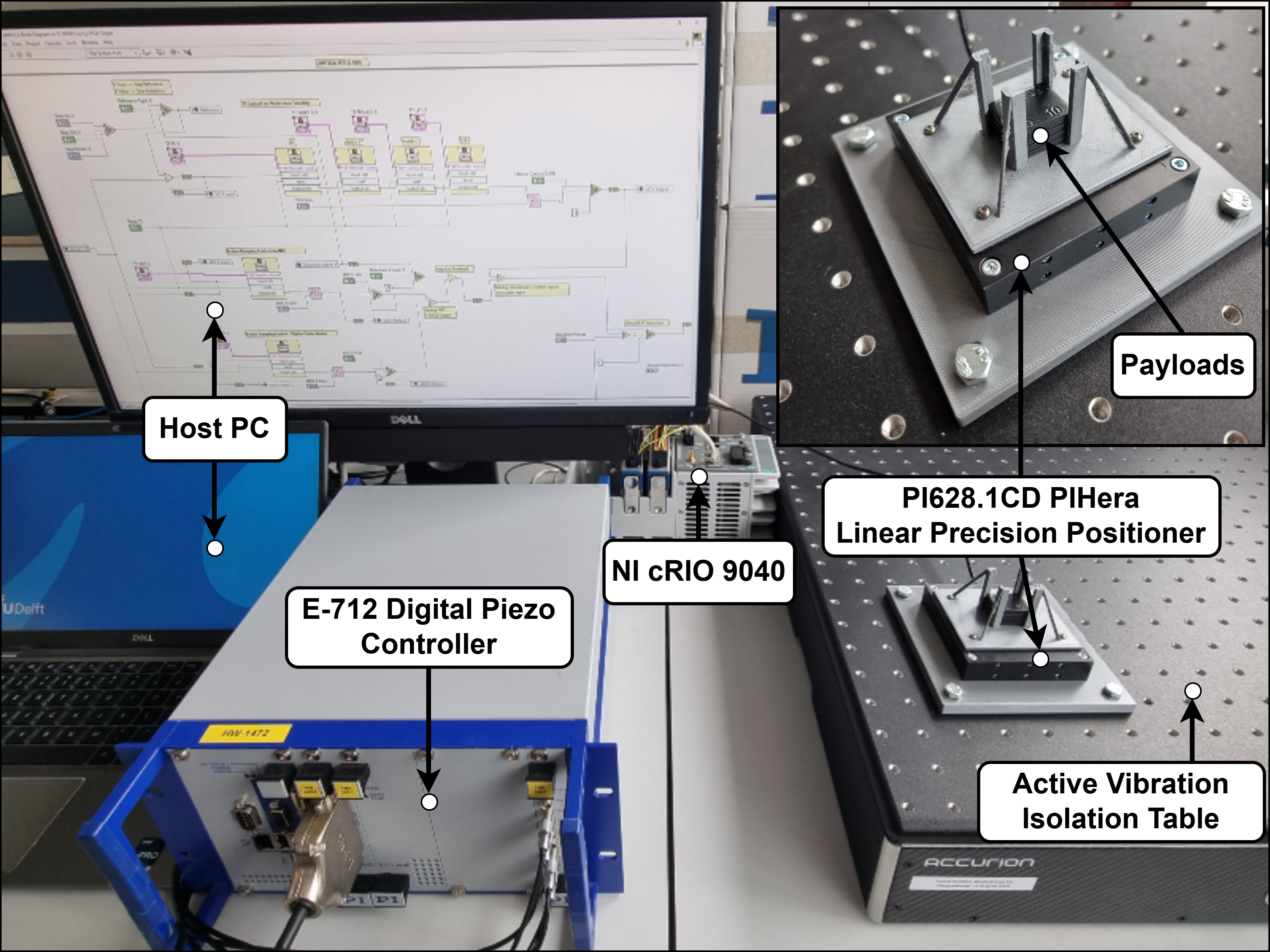}
\caption{}
\label{fig:ExperimentalSetup}
\end{subfigure}
\begin{subfigure}{\linewidth}
\centering
\includegraphics[width=0.9\linewidth]{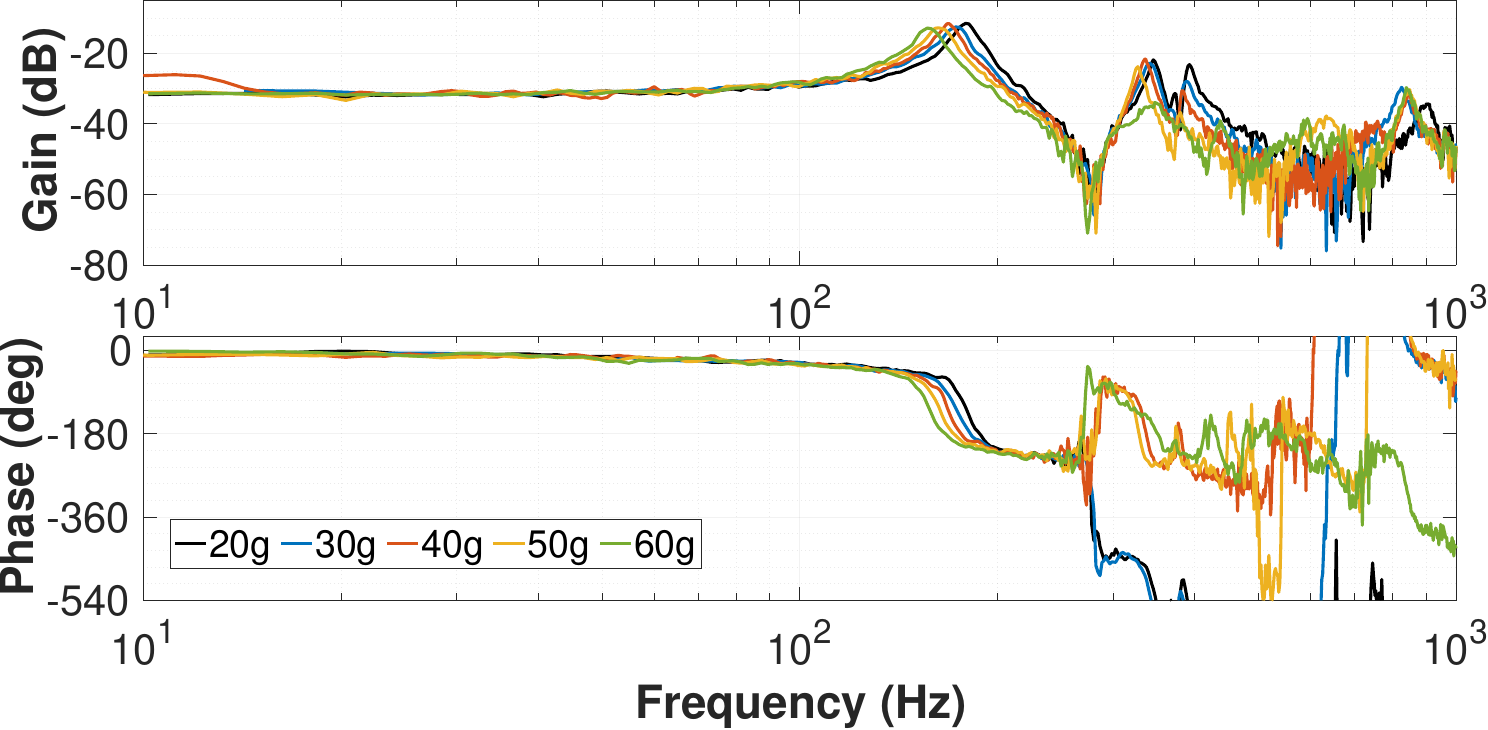}
\caption{}
\label{fig:SystemIdentification}
\end{subfigure}
\caption{(a) Experimental setup with P-628.1CD PIHera nanopositioner; (b) Identified system frequency responses under payload variations.}
\label{fig:setup_id}
\end{figure}

The system dynamics of a commercial P-628.1CD PI Hera nanopositioner are utilized. The single-axis stage incorporates a multilayer piezo-stack actuator, a flexure-guided mechanism for frictionless motion, and a capacitive displacement sensor for feedback. The stage is driven through a voltage amplifier and signal-conditioning modules in the E-712 controller. External control is implemented via NI CompactRIO with an embedded FPGA and 16-bit analog I/O, running a LabVIEW control algorithm. The actuation range is 0–10 V, with a sampling time of $t_s = 30~\mu s$. System dynamics were identified using a 0.1 V white-noise input (0.1 V offset, $0.0033~V^2$ variance). The sensor output was processed in MATLAB, where FRFs were estimated using Chebyshev windowing. Payload effects were characterized by adding masses in 10 g increments. 

The measured FRF exhibits four resonance modes up to 1000 Hz. The first resonance decreases from 179 Hz to 156 Hz with increasing payload, accompanied by similar downward shifts in higher modes, as shown in Fig.~\ref{fig:SystemIdentification}. Specifically, the second mode shifts from 264 Hz to 256 Hz, the third from 350 Hz to 326 Hz, and the fourth from 905 Hz to 840 Hz. Although the second mode is non-dominant with low amplification, it introduces a phase shift that is relevant for control design. Nevertheless, the low amplification at this resonance results in minimal variation in its characteristics. The phase lag observed below the first resonance is due to the actuator–amplifier dynamics and delay, and the pole–zero interlacing confirms the collocated configuration.

\subsection{Uncertainty Model}
\label{sec:unc_ex}
\begin{figure*}
\centering
\includegraphics[width=0.8\textwidth]{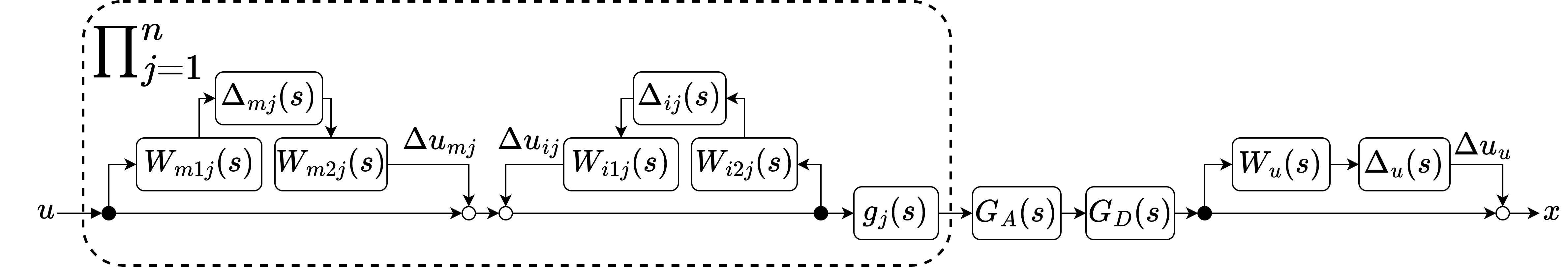}
\caption{Generalized form of the uncertainty model employed for payload-dependent nanopositioning system dynamics.}
\label{fig:unc_model_gen}
\end{figure*}
\begin{figure}[!t]
    \centering
        \includegraphics[width=0.8\linewidth,trim={1.7cm 3.7cm 1.7cm 3.6cm}, clip]{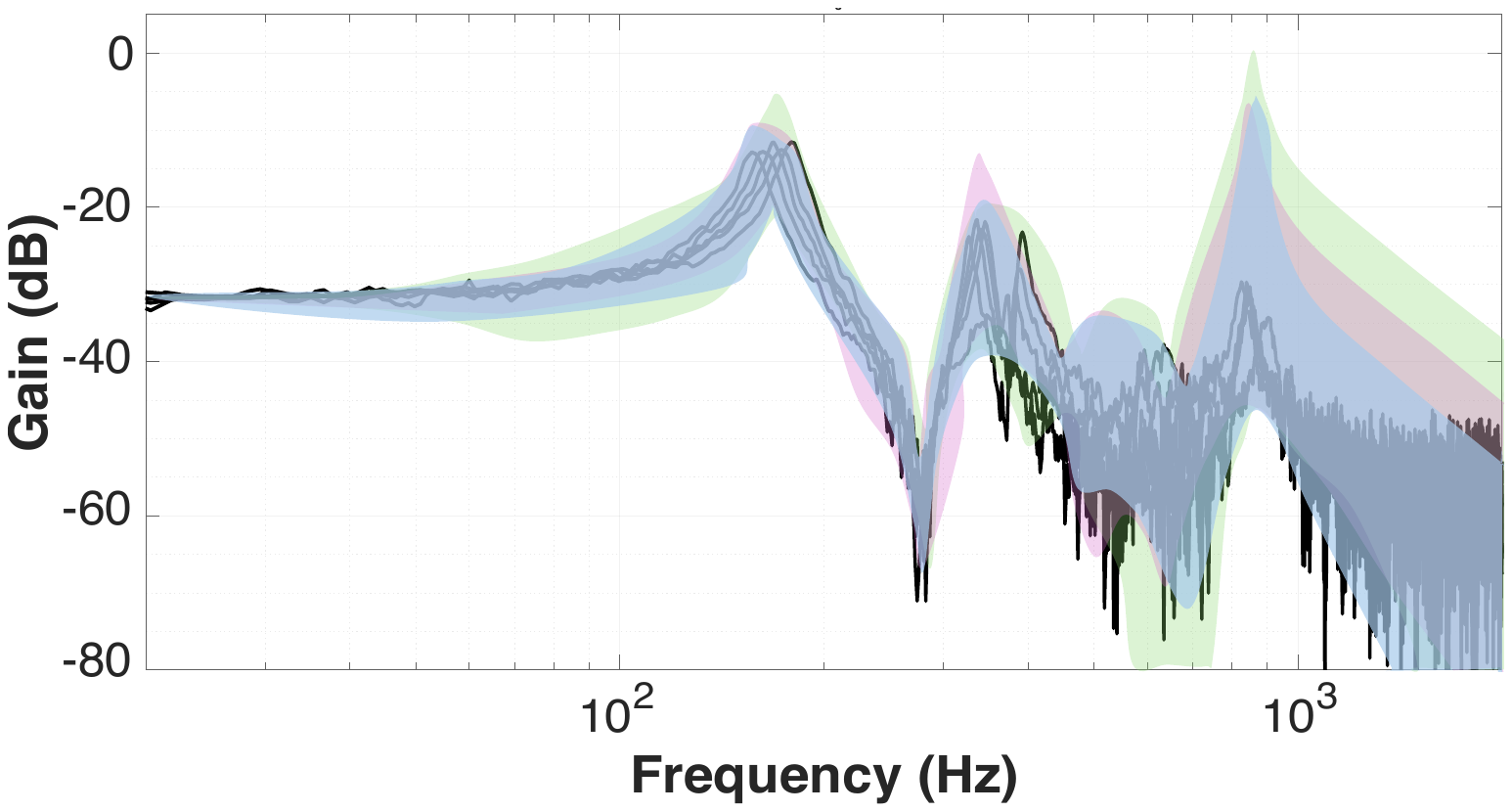}
    \caption{Uncertainty envelope for Model $M^{01}$: \colorbox[rgb]{0.886,0.953,0.851}{}, Model $M^{11}$: \colorbox[rgb]{0.937,0.831,0.933}{}, Model $M^{31}$: \colorbox[rgb]{0.780,0.851,0.941}{}, compared with the measured plant set (\textcolor{black}{\rule[0.5ex]{1em}{2.5pt}}).}
    \label{fig:unc_env}
\end{figure}
Four resonant modes are identified in the measured FRF up to 1000~Hz (as per Fig.~\ref{fig:SystemIdentification}), with payload variations shifting both resonant and anti-resonant frequencies. Beyond 1000~Hz, the reduced signal-to-noise ratio limits reliable identification. As discussed in Section~\ref{sec:uncert}, increasing the number of uncertainty blocks reduces conservatism but increases synthesis complexity. To explore this trade-off, three uncertainty models of increasing complexity are constructed. For these models, \( g_j(s) \) denotes the nominal transfer function of the \( j^{\text{th}} \) resonance and anti-resonance pair, while \( G_A(s) \) and \( G_D(s) \) represent the actuator dynamics and delay, respectively. The input multiplicative and inverse multiplicative uncertainty weights for the \( j^{\text{th}} \) pair, given by \eqref{eqn:weight_mult_invmult}, are \( W_{m1j}(s) \), \( W_{m2j}(s) \), and \( W_{i1j}(s) \), \( W_{i2j}(s) \) respectively, with corresponding uncertainty blocks \( \Delta_{mj}(s) \) and \( \Delta_{ij}(s) \) (as per \eqref{eqn:Delta_mult_invmult}). \(W_u(s) \) represents the unstructured output multiplicative uncertainty weight and $\Delta_u(s)$ is the corresponding uncertainty block. The general form of the uncertainty model is shown in Fig.~\ref{fig:unc_model_gen} and is described by:
\begin{equation}
\resizebox{1\hsize}{!}{$
G_p(s)=\left[\prod_{j=1}^n\mathbb{D}_{mj}(s)\cdot\mathbb{D}_{ij}(s)\cdot g_j(s)\right]\cdot G_A(s)\cdot G_D(s)\cdot\mathbb{D}_u(s);
$}
\end{equation}
where, $\mathbb{D}_{mj}(S)=1+W_{m1j}(s)\Delta_{mj}(s)W_{m2j}(s)$ is the structured input multiplicative uncertainty for the $j^{\text{th}}$ anti-resonance, $\mathbb{D}_{ij}(s)=\left(1-W_{i1j}(s)\Delta_{ij}(s)W_{i2j}(s)\right)^{-1}$ is the structured input inverse multiplicative uncertainty for the $j^{\text{th}}$ resonance, and $\mathbb{D}_u(s)=1+W_u(s)\Delta_u(s)$ is the unstructured output multiplicative uncertainty. For $j=1$, there is no anti-resonance. Therefore, we set $\Delta_{m1}=0$, $\implies\mathbb{D}_{m1}=1$. 

The three uncertainty models of increasing complexity are described below:
\begin{enumerate}
    \item Model $M^{01}$: a single unstructured uncertainty block representing variations in all resonant and anti-resonant modes, as well as high-frequency dynamic uncertainties. Hence, we set $\Delta_{mj}=\Delta_{ij}=0\;\text{for}\;j=1,2,3,4$.
    \item Model $M^{11}$: 1 structured uncertainty block for the dominant first resonance, with an unstructured uncertainty block accounting for variations in higher-order modes and high-frequency dynamic uncertainty. Therefore, we set $\Delta_{mj}=0\;\text{for}\;j=1,2,3,4$ and $\Delta_{ij}=0\;\text{for}\;j=2,3,4$.
    \item Model $M^{31}$: 3 structured uncertainty blocks corresponding to the dominant \(1^\text{st}\), \(3^\text{rd}\), and \(4^\text{th}\) resonant modes, with an unstructured uncertainty block capturing variations in the non-dominant \(2^\text{nd}\) mode and high-frequency dynamic uncertainty. Thus, we set $\Delta_{mj}=0\;\text{for}\;j=1,2$ and $\Delta_{ij}=0\;\text{for}\;j=2$.
\end{enumerate}

The envelopes of the uncertain plant models obtained using the aforementioned uncertainty modelling approaches are shown in Fig.~\ref{fig:unc_env}. As expected, the model employing a single unstructured uncertainty block yields the simplest formulation but also the most conservative representation of the measured plant set. Incorporating structured uncertainty blocks associated with individual system modes progressively reduces this conservatism.

\subsection{Controller Structure}\label{sec:BP}
A fixed-structure bandpass controller is synthesized using mixed-$\mu$ synthesis for each of the proposed uncertainty modelling approaches. The controller has the form:
\begin{equation}\label{eqn:BP}
    C(s) = M\left(\frac{s}{s^{2} + 2\zeta_{c}\omega_{c}s + \omega_{c}^{2}}\right)^{n}
      \left(\frac{s - \omega_{d}}{s + \omega_{d}}\right).
\end{equation}
This structure implements an $n^{\text{th}}$-order bandpass element that creates a bandgap at $\omega_{c}$. Since such a controller induces a phase drop of approximately $n\pi$, a non-minimum-phase filter with a corner frequency $\omega_d$ is included to smooth the phase near the bandgap and maintain adequate stability margins. The overall gain is set by the parameter $M$. The controller structure in \eqref{eqn:BP} provides damping within a targeted narrow frequency band while avoiding amplification of system dynamics elsewhere. Its narrow-band nature localizes any spillover effects arising from the damping action. Moreover, the low controller gain at both low and high frequencies attenuates the effect of high-frequency measurement noise (\cite{natu5947393narrow}).

The controller structure, initial parameter values, and parameter bounds are specified, and structured mixed-$\mu$ synthesis is used to compute the optimal controller parameters. Initial values and bounds are selected based on the resonant mode targeted for damping. For instance, the initial bandgap frequency ($\omega_c$) is set near the minima of the first resonance of the plant set, with bounds spanning the observed resonance variation. The damping ratio ($\zeta_c$) bounds are chosen to avoid values that are too low, which reduce robustness to resonance variations, or too high, which, while robust, may inadequately damp the resonance in closed-loop.

\subsection{Generalised Plant Formulation and Weight Design}
\label{sec:genp}
\begin{figure}[t!]
\centering
    \includegraphics[width=0.65\linewidth]{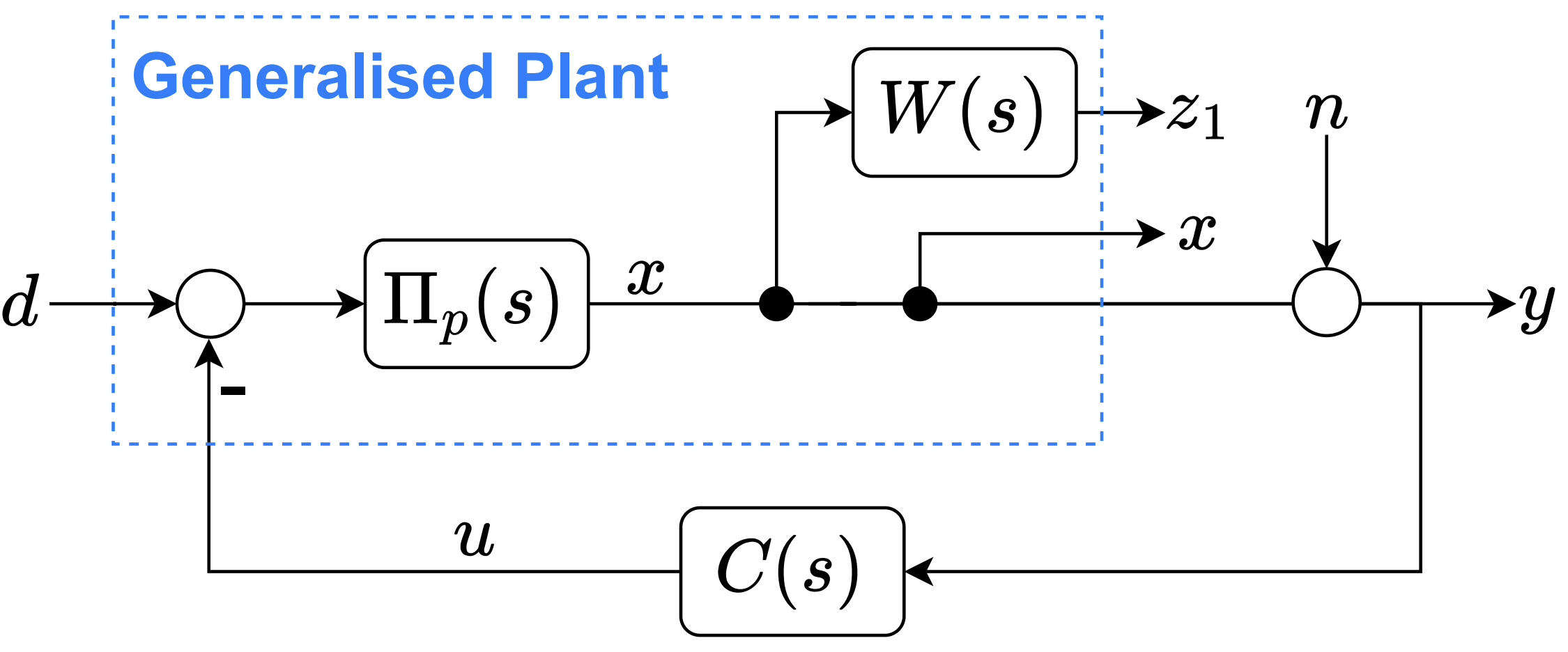}
\caption{General architecture with generalised plant for structured mixed-$\mu$ synthesis.}
\label{fig:genp_ADC}
\end{figure}
\begin{figure*}
\centering
\subfloat[\parbox{0.08\textwidth}{\centering{\scriptsize Model $M^{01}$}}]{\includegraphics[width=0.33\textwidth]{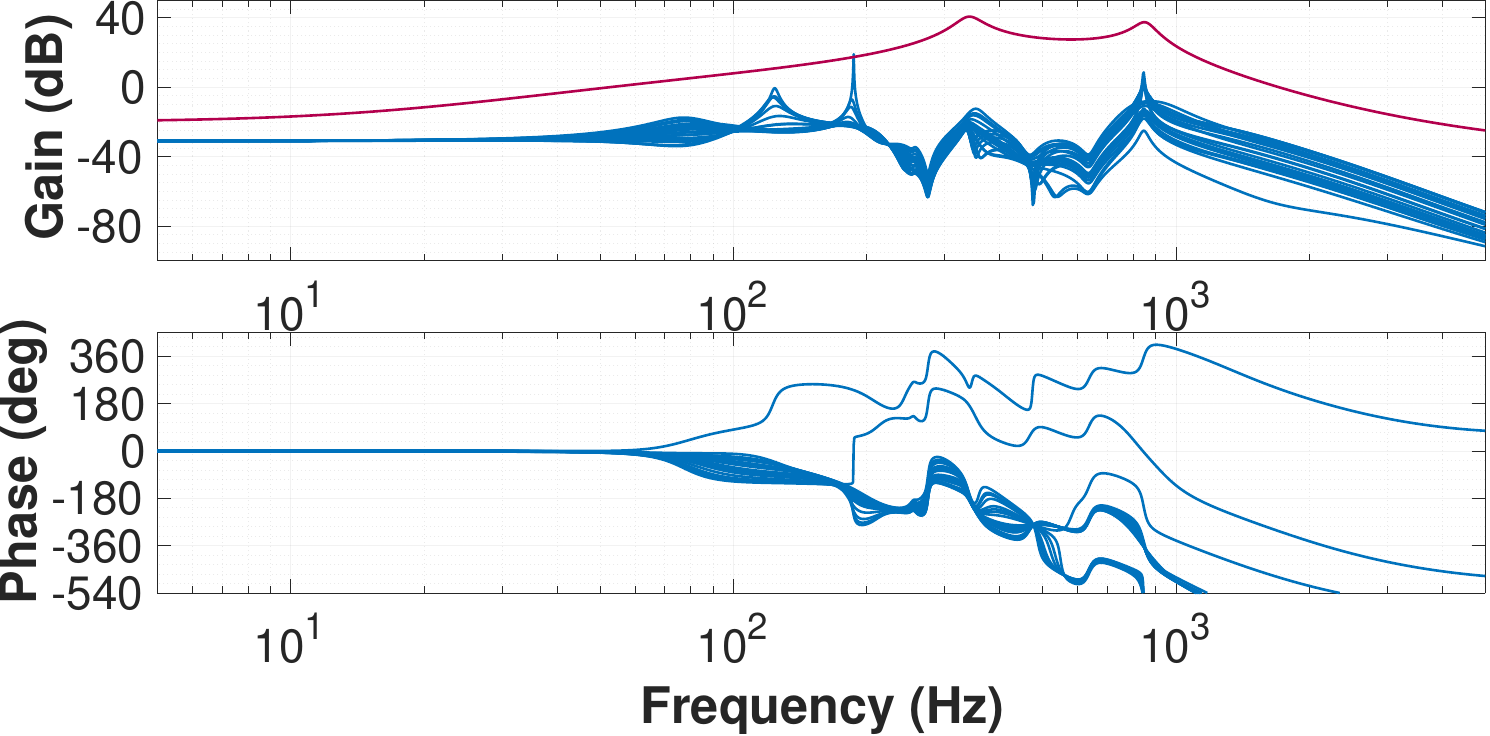}\label{unstr_ps01}}
\hfil
\subfloat[\parbox{0.08\textwidth}{\centering{\scriptsize Model $M^{11}$}}]{\includegraphics[width=0.33\textwidth]{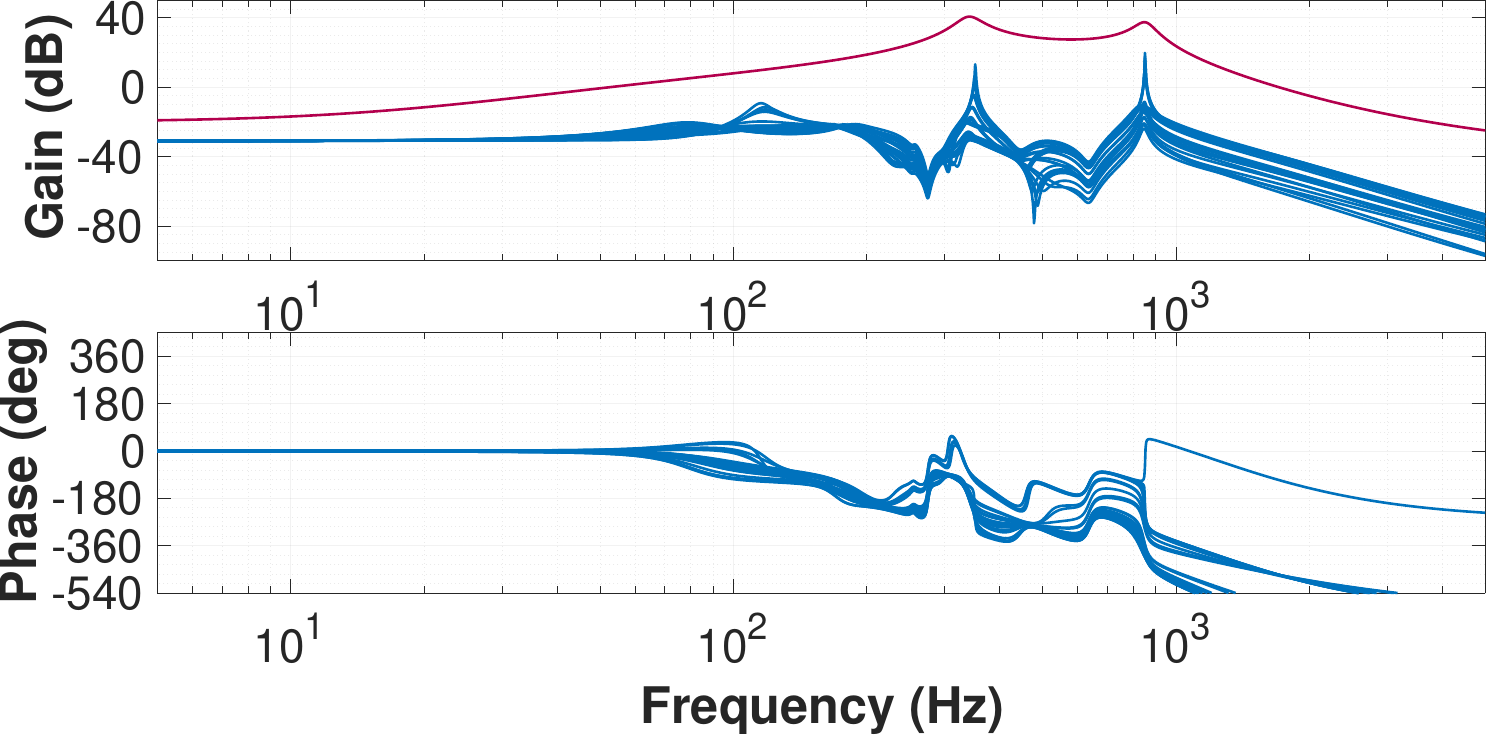}\label{1str1unstr_ps11}}
\hfil
\subfloat[\parbox{0.08\textwidth}{\centering{\scriptsize Model $M^{31}$}}]{\includegraphics[width=0.33\textwidth]{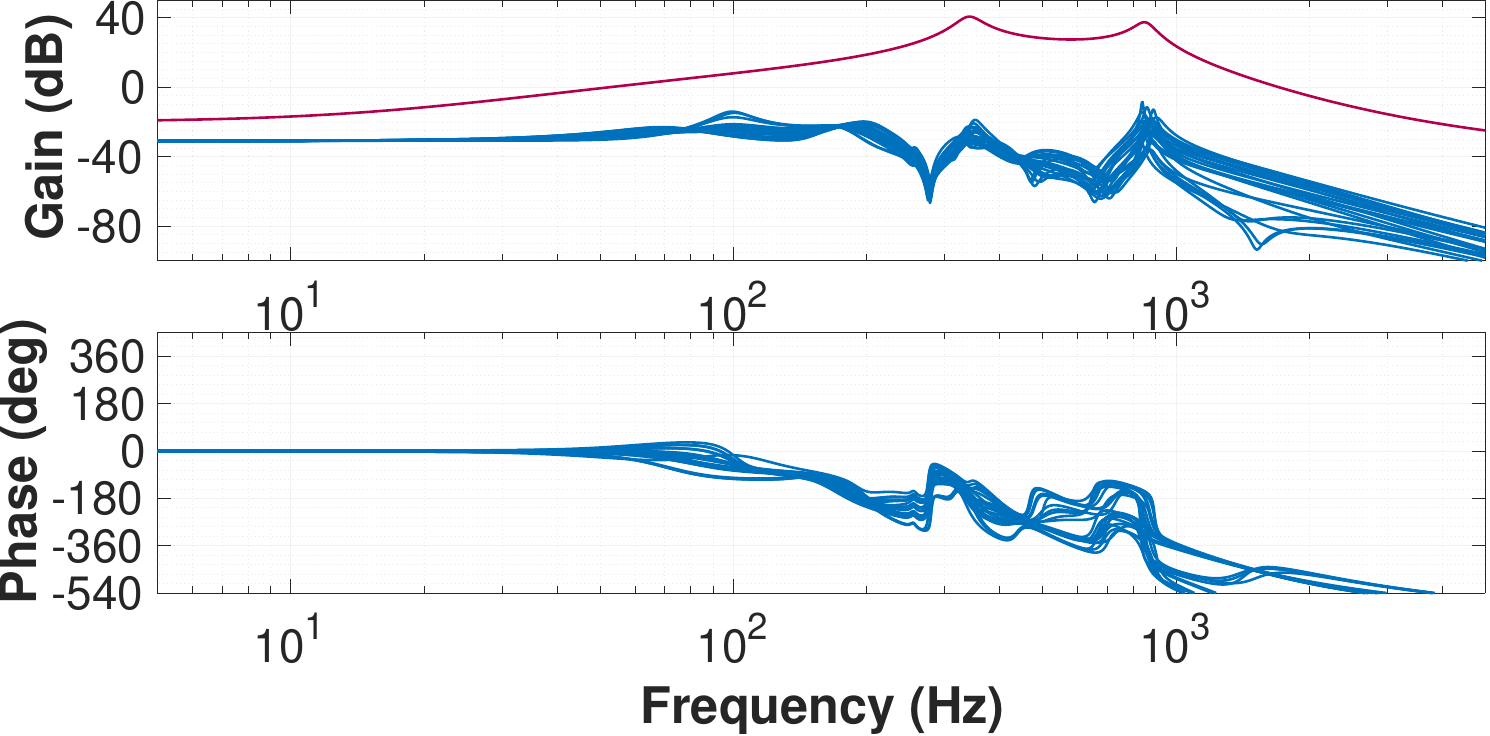}\label{3str1unstr_ps31}}
\hfil
\caption{Process sensitivity $\mathcal{PS}_{xd}^p(s)$ (\textcolor[rgb]{0 0.4470 0.7410}{\rule[0.5ex]{1em}{2.5pt}}) for the uncertain plant model with the inverse sensitivity weight $1/W(s)$ (\textcolor[rgb]{0.70,0.00,0.30}{\rule[0.5ex]{1em}{2.5pt}}).}
\label{fig:musyn_res}
\end{figure*}
\begin{figure*}
\centering
\subfloat[\parbox{0.4\textwidth}{\centering{}}]{\includegraphics[width=0.35\linewidth]{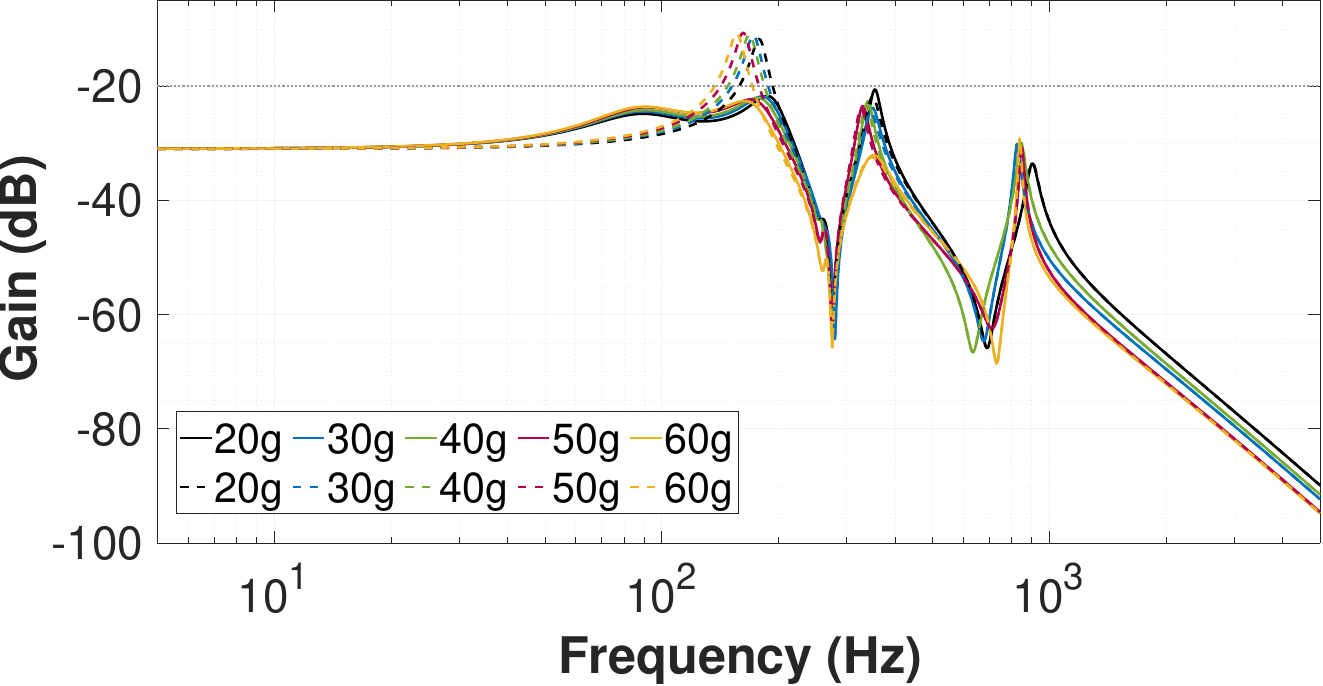}\label{fig:PS_G_comp}}
\hfil
\subfloat[\parbox{0.35\textwidth}{\centering{}}]{\includegraphics[width=0.35\linewidth]{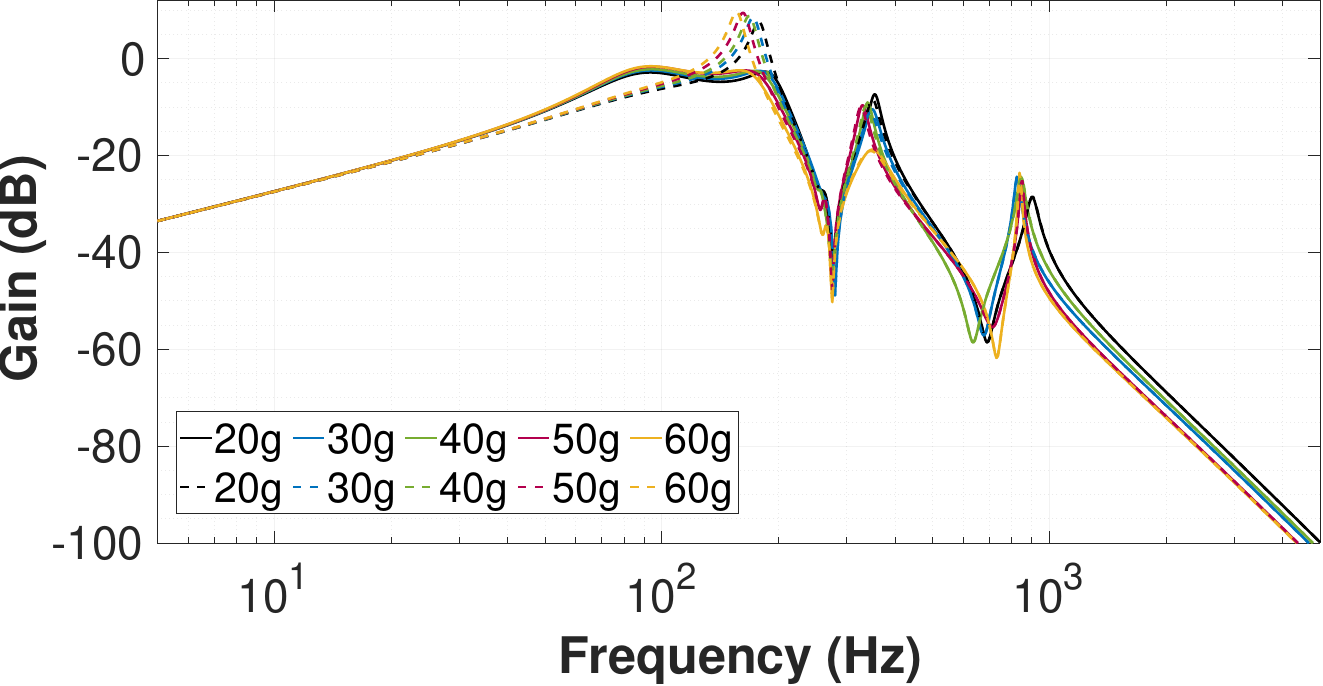}\label{fig:Sxn_L_comp}}
\hfil
\caption{(a) Comparison of process sensitivity $\mathcal{PS}_{xd}(s)$ (solid) with measured plant dynamics $G(s)$ (dashed) to evaluate damping performance; (b) Noise attenuation via $\mathcal{S}_{xn}(s)$ (solid) and stability assessment using $\mathcal{L}(s)$ (dashed).}
\label{fig:perf}
\end{figure*}
The architecture shown in Fig.~\ref{fig:genp_ADC} is employed for active damping and bandpass control. From an active damping standpoint, the process sensitivity $\mathcal{PS}_{xd}^p(s)$ associated with the uncertain plant model $G_p(s)$ must satisfy the following performance objective:
\begin{equation}
    \mathcal{PS}_{xd}^p(s) = \frac{G_p(s)}{1 + G_p(s)C(s)} \ll G_p(s)\:\text{for}\:\omega=\omega_c, \label{eqn:dist_rej}
\end{equation}
where $\omega_c$ is the bandgap frequency, corresponding to the first resonance of the system. The output \(x\) is regulated to meet the active damping requirement stated in \eqref{eqn:dist_rej}. The resulting generalized plant is:
\begin{equation}
    \begin{Bmatrix}
        z_1\\x
    \end{Bmatrix}=\underbrace{\begin{bmatrix}
        W(s)\cdot G_p(s) & -W(s)\cdot G_p(s)\\G_p(s) & -G_p(s)
    \end{bmatrix}}_{P(s)}\begin{Bmatrix}
        d\\u
    \end{Bmatrix}.
\end{equation}
Here, \( W(s) \) denotes the weighting function applied to regulate the output \( x \), shaping \( \mathcal{PS}_{xd}^p(s) \). 

The sensitivity weight exhibits a high-pass-filter-like shape with two notches near the dominant higher-order resonances. The low-frequency bound is chosen to limit the low-frequency magnitude of \( \mathcal{PS}_{xd}^p(s) \) below the maximum allowable amplification up to the bandgap frequency. The slope is selected to ensure that the resulting control design exhibits a high-frequency roll-off, guaranteeing attenuation of inputs beyond the relevant frequency range. A relaxation is introduced near the $3^{\text{rd}}$ and $4^{\text{th}}$ resonant modes, since the uncertainty models may overestimate the uncertainty at these higher-order modes, potentially resulting in amplified closed-loop responses that are unlikely to occur in practice. The degree of this relaxation is determined by Model $M^{01}$, as it represents the most conservative case among all studied uncertainty models. Overall, the weighting function balances low-frequency amplification constraints with high-frequency relaxation, localizing the spillover effects of damping at the first resonance. The same weight is utilized for all three uncertainty modelling methods during structured mixed-\(\mu\) synthesis.

\subsection{Structured Mixed-$\mu$ Synthesis}
\begin{table}[!b]
\centering
\caption{Structured mixed-$\mu$ synthesis results.}
\begin{tabular}{p{2cm} p{1.5cm} p{1.5cm} p{1.5cm} }
\hline
        Parameter & Model $M^{01}$ & Model $M^{11}$ & Model $M^{31}$\\\hline
        $\mu$ & \multicolumn{1}{c}{2.54} & \multicolumn{1}{c}{1.31} & \multicolumn{1}{c}{1}\\
        \multirow{2}{2cm}{Computation Time} & \multicolumn{1}{c}{8 s} & \multicolumn{1}{c}{490 s} & \multicolumn{1}{c}{3092 s} \\
        & & &\\
        \multirow{2}{2cm}{Gain reduction at $1^{\text{st}}$ mode} & \multicolumn{1}{c}{$\sim$10 dB} & \multicolumn{1}{c}{$\sim$10 dB} & \multicolumn{1}{c}{$\sim$10 dB}\\
        & & &\\\hline
    \end{tabular}
    \label{tab:musyn}
\end{table}

Structured mixed-$\mu$ synthesis was carried out to determine the optimal parameters of the bandpass controller in \eqref{eqn:BP} for the uncertainty models described in Section~\ref{sec:unc_ex}, using the generalized plant formulation and weight design from Section~\ref{sec:genp}. The synthesis was performed using MATLAB's \texttt{musyn} function in the \textit{Robust Control Toolbox}. All computations were executed on a machine with an 8-core Apple M2 3.5~GHz processor and 8~GB RAM.

Selecting a bandpass controller allows for targeted damping of the resonant mode while confining spillover to a localized frequency region. Furthermore, the low controller gain at low- and high-frequency aids in noise attenuation. Prescribing the controller structure reduces the computational complexity of the synthesis, offsetting the complexity introduced by the proposed uncertainty modelling approaches. A summary of the synthesis results is provided in Tab.~\ref{tab:musyn}, and the corresponding process sensitivity $\mathcal{PS}_{xd}^p(s)$ for the uncertain plant model is shown in Fig.~\ref{fig:musyn_res}.

As the uncertainty model becomes more detailed, the computation time increases considerably (see Tab.~\ref{tab:musyn}). Conversely, reducing computational complexity increases conservatism, causing the synthesis to consider an unnecessarily large plant set (see Fig.~\ref{fig:unc_env}). This results in control designs that exhibit amplified higher-order modal responses and, in some cases, instability over the modelled uncertainty set. This is highlighted
by the process sensitivity obtained using Model $M^{01}$ (Fig.~\ref{unstr_ps01}), which shows amplification not only at higher-order modes but also at low frequencies, leading to instability. The achieved $\mu=2.54$ indicates that the controller stabilizes only a fraction ($1/2.54$) of the modelled uncertainty, highlighting the excessive conservatism of Model $M^{01}$ in the entire frequency range of interest.

For Model $M^{11}$, the inclusion of a structured uncertainty block to account for variations in the first resonance significantly reduced the conservatism of the uncertainty model, improving the achieved $\mu$ to 1.31. However, the uncertainty model remains conservative at higher-order modes, where amplification is observed and extreme cases approach instability (see Fig.~\ref{1str1unstr_ps11}).

Model $M^{31}$ reduces conservatism across the entire frequency range. Consequently, synthesis using Model $M^{31}$ produces a bandpass controller that is stable across the entire uncertain plant model and meets the weight specifications, achieving $\mu = 1$ (see Fig.~\ref{3str1unstr_ps31}).

The controller synthesized using Model $M^{31}$ is implemented in closed-loop on the measured plant set. For closed-loop stability, the phase margin $\phi_m$ is evaluated at all 0~dB crossings of the open-loop transfer function $\mathcal{L}(s)=G(s)C(s)$. Typically, a phase margin of $\phi_m \geq 30\degree$ is required for robustness (\cite{natu2025robust}). As shown in Fig.~\ref{fig:Sxn_L_comp}, for the measured plant set, the first 0~dB crossover exhibits $\phi_m \in [90\degree, 113\degree]$, while the second 0~dB crossover shows $\phi_m \in [90\degree, 95\degree]$, ensuring robust stability. Approximately a 10~dB reduction in amplification is observed around the first resonance in the process sensitivity (see Fig.~\ref{fig:PS_G_comp}), indicating effective damping. As discussed in Section~\ref{sec:BP}, the bandpass controller in \eqref{eqn:BP} contributes to noise attenuation, as shown by the process sensitivity $\mathcal{S}_{xn}(s)$ in Fig.~\ref{fig:Sxn_L_comp}. $\mathcal{S}_{xn}(s)$ provides a measure of the impact of noise on the actual system output \(x\), with lower values indicating better noise performance. The results indicate that the synthesized bandpass controller maintains $\mathcal{S}_{xn}(s)$ well below 0~dB in the low- and high-frequency regions, providing substantial noise attenuation, while near the bandpass frequency, $\mathcal{S}_{xn}(s)$ remains close to, but still below 0~dB, allowing for limited noise attenuation performance.

\section{Conclusion}
\label{sec:conclusion}
This work presents an uncertainty modelling approach that employs multiple uncertainty blocks to reduce conservatism while capturing variations in plant dynamics. A mathematical framework was developed to construct structured and unstructured uncertainty weights corresponding to parametric and dynamic uncertainties. The methodology was applied to model payload-induced variations in an industrial piezoelectric nanopositioning system. Uncertainty models of varying complexity were formulated and utilized to synthesize a bandpass controller via structured mixed-$\mu$ synthesis, thereby enhancing damping performance. The resulting controllers were compared in terms of computation time and model conservatism. The study demonstrated that the proposed method provided a minimally conservative representation of large plant variations, encompassing both multi-parameter variations and dynamic uncertainties, compared to conventional single-block uncertainty formulations. Additionally, with these detailed uncertainty models, controllers that satisfy robust performance requirements can be synthesized, albeit at the cost of increased computational effort. 

Ongoing research focuses on applying the proposed uncertainty modelling methodology to robust control synthesis for systems with variations in multiple plant parameters and dynamic uncertainties in dual-loop configurations for active damping and motion control. The simulation results presented here will also be validated experimentally to evaluate practical performance and robustness under real operating conditions. Furthermore, the methodology can be extended to MIMO systems with direction-dependent or cross-coupled dynamic variations.

\begin{ack}
The authors express their sincere gratitude to Mathias Winter, Head of Piezo System \& Drive Technology, and Dr.-Ing. Simon Kapelke, Head of Piezo Fundamental Technology, Physik Instrumente (PI) SE \& Co. KG, for their invaluable collaboration in providing technical insights concerning the system and its applications.
\end{ack}

\bibliography{example}
\end{document}